# Shape-Optimized Electrooptic Beam Scanners: Experiment

J. C. Fang, M. J. Kawas, J. Zou, V. Gopalan, T. E. Schlesinger, *Senior Member, IEEE,* and D. D. Stancil, *Senior Member, IEEE*

*Abstract*—A new horn-shaped electrooptic scanner is described with significantly improved scanning sensitivity over rectangular-shaped devices. In the new device, the shape of the scanner is chosen to follow the trajectory of the beam. An example design is described that exhibits a factor of two larger scanning sensitivity than a rectangular device with comparable maximum scanning angle. Beam propagation simulations and measurements on an experimental device verify the scanner performance.

*Index Terms*—Electrooptic devices, electrooptic effects.

RECTANGULAR electrooptic beam scanners fabricated using the domain inversion method have demonstrated high-speed voltage-controlled beam steering [1]–[4]. Such devices have potential applications in optical data storage, displays, laser scanning/printing, and other integrated optical systems. An attractive feature of the domain-inversion fabrication method is that domains with virtually any shape can be readily created. In this paper we show how this capability is used to realize continuously tapered scanners with significantly improved performance compared to comparable rectangular devices.

Although easy to design and analyze, rectangular scanners have an important drawback. The width of such devices must be large enough to accommodate the full bipolar deflection of the beam at the exit. This restriction forces the scanner input dimension to be unnecessarily large and subsequently limits the deflection sensitivity. In a companion paper [6], analysis is presented showing how this drawback can be overcome by incorporating the trajectory of the deflected beam into the shape of the scanner, resulting in a new horn-shaped geometry. In this paper we report the first experimental horn-shaped scanner and compare its performance with simulations using the Beam Propagation Method (BPM).

The basic device concept is based on the fact that when an electric field is applied along the dielectric polarization direction of a LiTaO$_3$ crystal, there is a change of the extraordinary index of refraction $n_e$. The sign of the change depends upon the direction of the applied electric field with respect to the spontaneous direction of polarization and is given by [7]

$$\Delta n_e = \tfrac{1}{2} n_e^3 r_{33} E_3 \quad (1)$$

where $\Delta n_e$ is the change of index of refraction, $n_e$ is the original extraordinary index without the applied field, $r_{33}$ is the corresponding electrooptic coefficient of the structure, and $E_3$ is the applied electric field along the direction of polarization (denoted as subscript 3). When light crosses an interface between regions of oppositely oriented polarization, the total change of index across the interface is $\Delta n = 2\Delta n_e$.

For light traveling in a rectangular-shaped scanner, a small angle approximation for the angular deflection inside the crystal $\theta_{\text{int}}$ of the beam is given by [5], [6]

$$\theta_{\text{int}} = \frac{\Delta n}{n_e} \frac{L}{W} \quad (2)$$

where $L$ and $W$ are the length and width of the scanner, respectively. Upon exiting the scanner into air, Snell's law for small angles leads to a magnification of the deflection by the factor $n_e$. The deflection angle observed external to the crystal is, therefore,

$$\theta_{\text{ext}} = \Delta n \frac{L}{W}. \quad (3)$$

It follows that the width $W$ of the scanner should be small to ensure a high deflection sensitivity, but large enough to completely enclose the trajectory of the beam within the device.

From the above discussion, it is clear that the deflection sensitivity could be improved if the width was made as small as permitted by the beam diameter at the input, but was gradually increased to accommodate the trajectory of the beam. Assuming the length of the scanner is large relative to the width ($L \gg W$) and the maximum change in index is small compared to $n_e$ ($\Delta n \ll n_e$), the lateral position of the beam $x(z)$ and corresponding scanner width $W(z)$ are determined by [6]

$$\frac{d^2 x}{dz^2} = \frac{\Delta n}{n_e} \frac{1}{W(z)} \quad (4)$$

Manuscript received August 3, 1998; revised September 25, 1998. This work was supported in part by the National Science Foundation under grant ECD-8907068 and by the Data Storage Systems Center at Carnegie Mellon University.

J. C. Fang was with the Department of Electrical and Computer Engineering, Carnegie Mellon University, Pittsburgh, PA 15213 USA. She is now with the California Institute of Technology, Pasadena, CA USA.

M. J. Kawas was with the Department of Electrical and Computer Engineering, Carnegie Mellon University, Pittsburgh, PA 15213 USA. He is now with the U.S. Navy.

J. Zou, T. E. Schlesinger, and D. D. Stancil are with the Department of Electrical and Computer Engineering, Carnegie Mellon University, Pittsburgh, PA 15213 USA.

V. Gopalan was with the Department of Electrical and Computer Engineering, Carnegie Mellon University, Pittsburgh, PA 15213 USA. He is now with the Center for Material Science, Los Alamos National Laboratory, Los Alamos, NM 87545 USA.

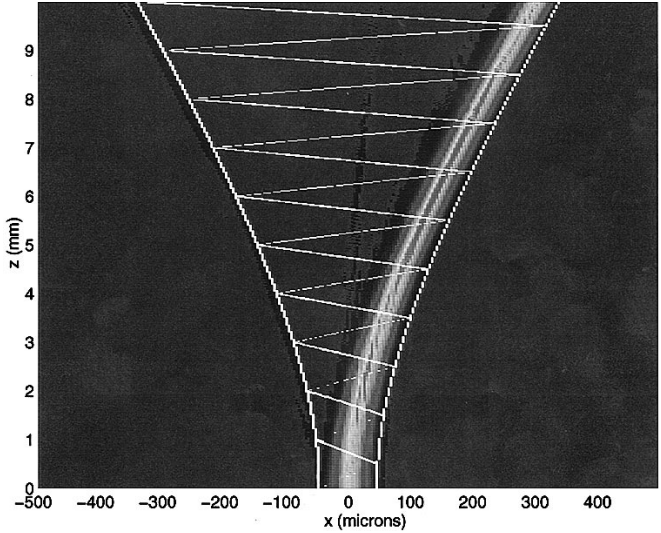

Fig. 1. BPM simulation of a light beam propagating through the horn-shaped device. The radius of the waist is 30 $\mu$m and is focused at the end of the scanner. The length of the device is 10 mm. (Note that the vertical and horizontal scales are different.)

$$\frac{W(z)}{2} = x(z) + \omega_0 \sqrt{1 + \left(\frac{\lambda_0(z-L)}{\pi n_e \omega_0^2}\right)^2} \quad (5)$$

where $\lambda_0$ is the free space wavelength, and $\omega_0$ is the beam waist. Equation (5) describes the change in gaussian beam diameter with position and assumes that the beam waist is placed at the output of the scanner. Using this theory, the scanner shown in Fig. 1 was designed.

The performance of the scanner shown in Fig. 1 was simulated using the beam propagation method [2]. The entrance of the horn-shaped scanner (shown at the bottom of Fig. 1) is 92 $\mu$m wide and the exit width is 678 $\mu$m. The extraordinary index and the corresponding electrooptic coefficient for LiTaO$_3$ for $\lambda_0 = 0.6328$ $\mu$m are $n_e = 2.1807$ and $r_{33} = 30.5$ pm/V, respectively. The change of index of refraction is $\Delta n = 2.1 \times 10^{-3}$ for an applied voltage of 1 kV across a substrate thickness of $d = 150$ $\mu$m. This voltage results in less than one third of the poling field of about 21 kV/mm, ensuring stable device operation. For the simulation, the waist of the input Gaussian beam was chosen to be 30 $\mu$m and placed at the exit of the device. The number of interfaces was chosen to be 20. To minimize the lateral truncation of the optical beam, the width of the simulated scanner was chosen to be 30% larger than that calculated from (4) to (5).

The predicted maximum external bipolar deflection angle of the horn-shaped device is 174.6 mrad (87.3 mrad on each side), which leads to a total of 13 resolvable spots (based on a separation by a $1/e^2$ beam diameter), including the center spot. The corresponding internal deflection angle is $\pm 40.0$ mrad. This performance is obtained with an applied voltage of $\pm 1$ kV. For comparison, consider a rectangular scanner capable of the same maximum deflection. Since the beam at the output of the scanner (but still inside the crystal) appears to pivot about the point at the center of the scanner [6], for small angles the maximum internal scanning angle imposed by geometrical constraints in a rectangular device (see Fig. 2)

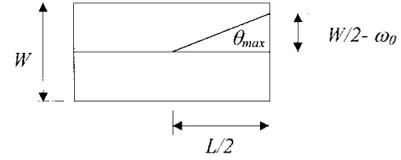

Fig. 2. Geometry showing maximum deflection angle for a rectangular scanner.

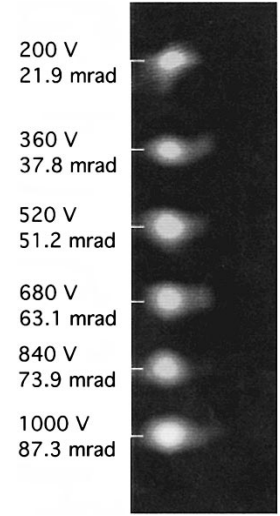

Fig. 3. A multiple exposure photograph showing the number of spots deflected by the horn-shaped scanner with 0.2–1 kV applied. Although not shown, similar operation extends to 0 V.

is approximately given by

$$\theta_{\max} = \frac{W - 2\omega_0}{L}. \quad (6)$$

Taking $\omega_0 = 30$ $\mu$m, $L = 10$ mm, and $\theta_{\max} = 40.0$ mrad, the required width of the rectangular scanner is 460 $\mu$m. From (2), the required value of $\Delta n$ to achieve this angle would be $4 \times 10^{-3}$, or about 2 times larger than that of the horn-shaped device. Consequently, the rectangular device would require twice the voltage of the horn-shaped device to achieve the same maximum scanning angle.

The optimized horn-shaped pattern described above was photolithographically transferred to a SAW grade $z$-cut LiTaO$_3$ substrate with thickness of 150 $\mu$m. The device fabrication followed the procedures described in [3] and [8].

A linearly polarized He–Ne laser ($\lambda = 0.6328$ $\mu$m) with a 750-$\mu$m diameter beam was used for testing. The polarization was adjusted to be parallel to the $z$ axis of the crystal. Since the optimum beam waist for the design is 30 $\mu$m, a cylindrical lens followed by a spherical lens was used to achieve the desired beam diameter at the exit of the scanner without truncation of the beam by the substrate.

The external deflection of the scanner was measured to be 91 mrad at 1 kV, in good agreement with the design value of 87.3 mrad. Fig. 3 shows a multiple exposure photograph of the deflected spots. The figure shows six resolvable spots with an applied voltage from 0.2–1 kV (the center spot at 0 V is omitted), corresponding to 13 total resolvable spots with bipolar deflection, again in agreement with the calculation. The spots are separated by their $1/e^2$ diameter.

In conclusion, we have demonstrated a novel horn-shaped electrooptic beam scanner. The deflection angle was measured to be 91 mrad corresponding to seven resolvable spots when the applied voltage ranged from 0 to 1 kV. The horn-shaped device exhibits superior deflection sensitivity compared to a rectangular device with the same length and total deflection. The improvement is most pronounced when large deflection angles are required.